\documentclass[preprint,showpacs,aps,preprintnumbers,amsmath,amssymb,
nofootinbib]{revtex4}
\usepackage{epsfig}
\begin{document}

\begin{flushright}
%\preprint{hep-ph/xxyyzz}
%\today\\
%UH-0528
\end{flushright}

%%%%%%%%%%%%%%%%%%%%%%%%%%%%%%%%%%%%%%%%%%%%%%%%%%%%%%%%%%%%%%%%%%%%%%%%

\newcommand{\be}{\begin{equation}}
\newcommand{\ee}{\end{equation}}
\newcommand{\bea}{\begin{eqnarray}}
\newcommand{\eea}{\end{eqnarray}}
\newcommand{\bers}{\begin{eqnarray*}}
\newcommand{\eers}{\end{eqnarray*}}
\newcommand{\nn}{\nonumber}
\newcommand\un{\cal{U}}
\def\dis{\displaystyle}

%%%%%%%%%%%%%%%%%%%%%%%%%%%%%%%%%%%%%%%%%%%%%%%%%%%%%%%%%%%%%%%%%%%%%%%

\title{\large Signature of new physics in $B \to \phi \pi $
decay }
\author{B. Mawlong$^1$, R. Mohanta$^1$  and A. K. Giri$^2$  }
\affiliation{$^1$ School of Physics, University of Hyderabad,
Hyderabad - 500 046, India\\
$^2$ Department of Physics, Punjabi University, Patiala - 147 002,
India }
%\vspace*{1.0 truein}

\begin{abstract}

We investigate  the effect of an  extra fourth quark generation and
FCNC mediated $Z$ and $Z'$ bosons on the rare decay mode $B^- \to
\phi \pi^-$. In the standard model, this mode receives only $b \to
d$ penguin contributions and therefore, highly suppressed with
branching ratio $\sim 5 \times 10^{-9}$. This in turn makes this
mode a very sensitive probe for new physics. We find that due to the
above mentioned  new physics contributions there is a significant
enhancement in its branching ratio. Furthermore, the direct
CP violation parameter which is identically zero in the SM
 is found to be quite significant. If this mode will be observed in
the upcoming LHCb experiment, it will not only provide a clear
signal of new physics but also  can  be used to constrain the new
physics parameter space.
\end{abstract}
\pacs{ 13.25.Hw, 12.60.-i, 12.60.Cn}
 \maketitle

The $B$ decay modes provide valuable insights to critically test
the standard model (SM) and to look for the possible existence
of physics beyond the SM.  One way of searching for new physics (NP) is by
studying the rare decay modes which are  induced by
flavor changing neutral current (FCNC) transitions. In the SM,
such rare decays arise at the one-loop level and thus the
study of the same will provide us an
excellent testing ground for NP.

Over the years, there has been profound interest in the search for
physics beyond the SM. The observed discrepancy between the measured
$S_{\phi K_S}$ and $S_{\psi K_S}$ \cite{S} already gave an
indication of the possible existence of NP in the $B \to \phi K_S$
decay amplitude and this has, in one way, motivated many to carry out
intensive search for NP. Although the presence of NP in the
$b$-sector is not yet firmly established, but there exist several
smoking gun signals \cite{soni} which will be verified in the
upcoming LHCb experiment or super B factories. Therefore, it is interesting to
examine as many different rare decay channels as possible to have an
indication of new physics.

In this paper, we would like to explore the effect of the extra fourth
generation of quarks and FCNC mediated $Z (Z')$ boson(s) in the rare
decay mode $B^- \to \phi \pi^-$, which is a pure penguin induced
process, mediated by the quark level transition $ b \to d  \bar s s$.
The interesting feature of this  process is that it is dominated by
the electroweak penguin contributions as the QCD penguins are OZI
suppressed, and  therefore  expected to be highly suppressed in the
SM. It, therefore, serves as a suitable place to search for new
physics. At present only the upper limit of its branching ratio
is known \cite{prd74}
\begin{equation}
{\rm Br}(B^{-} \to \phi \pi^{-}) < 0.24 \times 10^{-6} \;. \label{exp}
\end{equation}
 This decay mode has been analyzed both in the SM \cite{phism} and
 in various extensions of it \cite{phi} where  it has been
 found that in some of these new
physics models the branching ratio can be enhanced significantly
from its corresponding SM value.

In order  to discuss the effect of fourth quark generation and FCNC
mediated $Z$($Z'$) boson, we would first like to  present the SM
result using the QCD factorization \cite{qcd}. As the decay mode $B^- \to \phi
\pi^-$ proceeds through the quark level transition $ b \to
d  \bar s s$ and is a pure penguin induced process occurring at the
one loop level, the relevant effective Hamiltonian describing this
process is given by
\begin{equation}
{\cal H}_{eff}^{SM} = \frac{G_F}{\sqrt{2}} V_{pb} V_{pd}^*
\sum_{i=3}^{10} C_i(\mu) O_i\;,
\end{equation}
where $p = u, c$, $C_i(\mu)$'s are the Wilson coefficients evaluated
at the $b$-quark mass scale and  $O_i$'s are the QCD and electroweak
penguin operators.

In QCD factorization \cite{qcd}, the decay amplitude can be
represented in the form
\begin{equation}
A(B^- (p_B) \to \phi (\epsilon, p_1) \pi^-(p_2))=-i \frac{G_F}{\sqrt
2} 2 m_\phi f_\phi (\epsilon^* \cdot p_B) F_+^{B \pi}(0)\sum_{p=u,c}
\lambda_p (\alpha_3^p - \frac{1}{2} \alpha_{3,EW}^p) \;,
\end{equation}
where $ \lambda_p=V_{pb} V_{pd}^*$,  the QCD coefficients
$\alpha_{3(3,EW)}^p$ are related to the Wilson coefficients as
defined in \cite{qcd} and $F_+^{B \pi}$ is the form factor describing $B
\to \pi$ transition. It should be noted that the QCD coefficients
contributing to $B^- \to \phi \pi^-$ are independent of $p=u,c$,
(i.e., the virtual particles in the loop). Therefore, one can also
represent the above amplitude using CKM unitarity $\lambda_u
+\lambda_c+\lambda_t=0$,  as
 \begin{equation}
A(B^- \to \phi \pi^-)=i \frac{G_F}{\sqrt 2} 2 m_\phi f_\phi
(\epsilon^* \cdot p_B) F_+^{B \pi}(0) \lambda_t (\alpha_3 -
\frac{1}{2} \alpha_{3,EW}) \;,\label{amp}
\end{equation}
where we have now omited the superscripts on $\alpha$'s.
The above amplitude can be simplified by replacing $2 m_\phi
\epsilon^* \cdot p_B \rightarrow m_B^2$. The branching ratio thus
can  be obtained using the formula
\begin{equation}
{\rm Br}(B^- \to \phi \pi^-) = \frac{\tau_B}{16 \pi m_{B}} |A (B^- \to
\phi \pi^-) |^2 \;,
\end{equation}
where  $\tau_B$ is the lifetime of $B^-$ meson.
Another possible observable in this decay mode is the direct CP
violation parameter, defined as
\be
A_{CP}= \frac{\Gamma(B^+ \to \phi \pi^+)-\Gamma(B^- \to \phi \pi^-)}
{\Gamma(B^+ \to \phi \pi^+)+\Gamma(B^- \to \phi \pi^-)}\;.\label{acp}
\ee
In order to have non-zero direct CP violation, it is necessary that
the corresponding decay amplitude should contain at least two
interfering contributions with different strong and weak phases.
Since in the SM, this decay mode does not have two such different
contributions in its amplitude, the direct CP violation turns out to be
identically zero.

For the numerical
evaluation, we use the input parameters as given in the S4 scenario
of QCD factorization approach \cite{qcd}. The particle masses and lifetime of
the $B$ meson are taken from \cite{pdg}. The value of the form
factor at zero recoil is taken as $F_+^{B \pi}(0)$=0.28.
The value of the CKM matrix elements used are \cite{pdg},
$|V_{ub}|=3.96 \times
10^{-3}$, $|V_{ud}|=0.97383$, $|V_{cb}|=42.21 \times 10^{-3}$,
$|V_{cd}|=0.2271$ and $\gamma$ the phase associated with $V_{ub}$
as $70^{\circ}$. With these values as input parameters, the
branching ratio  obtained in the SM is
\begin{equation}
{\rm Br}^{SM}(B^- \to \phi \pi^-) = 4.45 \times 10^{-9} \;,
\end{equation}
which is quite below the experimental upper
limit as given in Eq. (\ref{exp}).

Now, in the presence of NP, the transition amplitude (\ref{amp})
receives additional contribution and can be symbolically represented
as \bea A^T(B^- \to \phi \pi^-) = A^{SM}+A^{NP}=A^{SM}(1+r~ e^{i
\delta}~ e^{-i(\beta-\phi)})\;, \eea where $\beta $ is the weak
phase of the SM amplitude i.e., we have used $V_{td}=|V_{td}| e^{-i
\beta}$ with value $\beta=0.375$, $\phi$ is the weak phase
associated with the NP amplitude and $\delta$ is the relative strong
phase between these two amplitudes. It should be noted that the
strong phases are generated by the final state interactions (FSI)
and at the quark level they arise through absorptive parts of the
perturbative penguin diagrams. Furthermore, $ r$ denotes the
magnitude of the ratio of NP to SM amplitude. Thus, we obtain the CP
averaged branching ratio $\langle {\rm Br} \rangle \equiv [{\rm
Br}(B^- \to \phi \pi^-)+{\rm Br} (B^+ \to \phi \pi^+ )]/2$,
including the new physics contribution  as \be \langle {\rm Br}
\rangle = {\rm Br}^{\rm SM} (1+r^2 +2 r \cos (\beta -\phi) \cos
\delta)\;,\label{br} \ee where $ {\rm Br}^{\rm SM} $ is the SM
branching ratio. It can be seen from the above equation that if $r$
is sizable, the branching ratio could be significantly enhanced from
its SM value in the presence of new physics. The direct CP violation
parameter (\ref{acp}) in the presence of NP becomes \be
A_{CP}=\frac{2 r \sin \delta \sin(\phi - \beta)}{1+r^2+2 r \cos
\delta \cos (\phi - \beta)}\;.\label{cp} \ee

We now consider the  effect of a sequential fourth generation of quarks
\cite{4gen}. This model is an extension of the SM with the addition
of a fourth quark generation. It retains all the features of the SM
except that it brings into existence the new members denoted by
$(t',b')$. The fourth up-type quark ($t'$) like $u, c, t$ quarks
contributes in the $b \to d$ transition at the loop level and hence
will modify the SM result. The effect of fourth generation of quarks in
various $B$ decays are extensively studied in the literature
\cite{4thgen,hou2}.

Due to the additional fourth generation, there will be mixing among
the new $b'$ quark and the three down type quarks of the SM and the
resulting mixing matrix will be a $4\times 4$ matrix. Accordingly
the  unitarity condition becomes $\lambda_u +\lambda_c + \lambda_t
+\lambda_{t'}=0$ and thus the effective Hamiltonian modifies as
\begin{equation}
{\cal H}_{eff} = -{\frac{G_F}{\sqrt{2}}} \biggr [ V_{tb} V_{td}^*
\sum C_i  O_i + V_{t'b} V_{t'd}^* \sum C_i^{t'}O_i \biggr],
\end{equation}
where $C_i^{t'}$ are the new Wilson coefficients arising due to the
$t'$ quark in the loop. The values of these Wilson coefficients at
the $M_W$ scale can be obtained from the corresponding contributions
from the $t$ quark by replacing the mass of $t$ quark in the Inami
Lim functions \cite{lim} by $t'$ mass (here we neglect the RG
evolution of these coefficients from $t'$ mass scale to the weak
scale $M_W$). These values can then be evolved to the $m_b$ scale
using renormalization group (RG) equation \cite{wilson}, as \be \vec
C_i(m_b)=U_5(m_b, M_W, \alpha ) \vec C(M_W)\;, \ee where $\vec C$ is
the $10 \times 1$ column vector of the Wilson coefficients and $U_5$
is the five flavor $10 \times 10$ evolution matrix. The explicit
forms of $\vec C(M_W)$ and $U_5(m_b, M_W, \alpha)$ are given in
\cite{wilson}. In Table-1, we present the values of these new Wilson
coefficients at $m_b$ scale for a representative set of values for
$m_{t'}=400$ GeV.
\begin{table}
\caption{ Values of the new Wilson coefficients at $m_b$ scale where
$C_i^{new}$ represents $C_i^{t'}$ for the fourth quark generation
model and $\tilde C_i$ for the FCNC mediated $Z$ boson model. The
phase  $\phi'=(\phi-\beta)$ is the relative weak phase between the
NP and SM amplitudes.}
\begin{tabular}{|c|ccc|ccc|ccc|}
\hline Wilson   &&& 4-Generation &&&Z boson model
&&& $Z'$ model\\
Coefficients   &&& ($m_{t'}$=400 GeV) &&&
$\left (\kappa=
\frac{|U_{bd}|}{|V_{tb} V_{td}^*|}\right )$&&& ($\xi_{L,R}=
|\xi_{L,R}|e^{i \phi'}$)\\
\hline
$ C_3^{new}(m_b) $ &&& 0.0195 &&& $0.19~ \kappa e^{i \phi'}$
&&& $0.05 ~\xi_L-0.01 ~\xi_R $\\
$ C_4^{new}(m_b) $ &&& $-0.0373$ &&& $-0.066~ \kappa
e^{i \phi'}$ &&& $-0.14 ~\xi_L+0.008 ~\xi_R $ \\
$ C_5^{new}(m_b) $ &&& 0.0101 &&& $0.009~ \kappa e^{i \phi'}$
&&& $0.029 ~\xi_L+0.017 ~\xi_R $ \\
$ C_6^{new}(m_b) $  &&& $-0.0435$ &&& $-0.031~ \kappa e^{i \phi'}$
 &&& $-0.162 ~\xi_L+0.01 ~\xi_R $ \\
$ C_7^{new}(m_b) $ &&& 0.0044 &&&$0.145~ \kappa e^{i \phi'}$
&&& $0.036 ~\xi_L-3.65 ~\xi_R $ \\
$ C_8^{new}(m_b) $ &&& $0.002$ &&&$0.053~ \kappa e^{i \phi'}$
&&& $0.01 ~\xi_L-1.33 ~\xi_R $ \\
$ C_9^{new}(m_b) $  &&& $-0.029$ &&& $-0.566~ \kappa e^{i \phi'}$
&&& $-4.41 ~\xi_L+0.04 ~\xi_R $ \\
$ C_{10}^{new}(m_b) $ &&& 0.0062 &&& $0.127~ \kappa e^{i \phi'}$
&&& $0.99 ~\xi_L-0.005 ~\xi_R $\\
\hline
\end{tabular}
\end{table}

After obtaining the values of the new Wilson coefficients at the $b$
quark mass scale, one can directly write the decay amplitude
analogous to (\ref{amp}), due to the fourth generation quarks  as
\be A^{NP} = {\frac{G_F}{\sqrt{2}}} 2 m_\phi f_\phi (\epsilon^* \cdot
p_B) F_+^{B \pi}(0) \lambda_{t'} \left (\alpha_3^\prime -
\frac{1}{2} \alpha_{3,EW}^\prime \right )\;.
\end{equation}
where  $\alpha'_{3(3,EW)}$'s are the new contributions arising from
the ${t'}$ quark contribution. We parameterize  the new CKM elements
as $\lambda_{t'} = r_{d} e^{i \phi}$, where $ \phi$ is the new weak
phase associated with $\lambda_t'$. Furthermore, since the unitarity
condition has now become modified the elements of the $3 \times
3$ upper submatrix of the $4 \times 4$ quark mixing matrix  will be
different from the corresponding values of SM  CKM matrix elements.
Since $V_{tb}$ and $V_{td}$ are not precisely known (i.e., not
directly extracted from the experimental data, but fitted using the
unitarity constraint) we will use the lower limits from \cite{pdg}
i.e., $|V_{tb}|=0.78$ and $|V_{td}|=7.4 \times 10^{-3}$.

In order to study the effect of the fourth generation,  we need
to know the
values of the new parameters $(m_{t'}, r_d, \phi)$. Based on an
integrated luminosity of $2.3 fb^{-1}$ CDF collaboration
\cite{cdf} gives the lower bound on
$m_{t'}$ as $m_{t'}>284$ GeV. Recently it has been shown
that  the observed pattern of deviations in the CP
symmetries of $B$ system can be explained in the fourth quark
generation model if $m_{t'}>700$ GeV \cite{soni08}. 
Therefore, in our analysis we
consider three representative values for $m_{t'}=400,~ 600$ and 800
GeV.  The value of $r_d$ can be obtained from the measured mass
difference $\Delta M_{B_d}$ of $B^0 - \bar B^0$ system and the
corresponding expression for $\Delta M_{B_d}$ in the presence of
fourth quark generation can be found in Ref. \cite{hou2}. Thus,  we
obtain the values $r_d$  for different $m_t'$,  consistent
with the unitarity condition of $4\times 4$ matrix as: $r_d \sim
-3.8\times 10^{-3}~(m_{t'}=400$ GeV),
 $r_d \sim -2.7\times 10^{-3}~(m_{t'}=600$ GeV)
and $r_d \sim -2.1\times 10^{-3}~(m_{t'}=800$ GeV).
Using these values, in  Figure-1 we show the variation
of the branching (left panel) and the direct CP asymmetry (right
panel) with the new weak phase $\phi$ for three different values of
$m_{t'}$. From the figure one can see that the branching ratio is
significantly enhanced from its SM value and this enhancement is
 more pronounced for large $m_{t'}$.  It should also be noted that
 non-zero direct CP violation in this mode could
be possible in the presence of an additional generation
of quarks.

\begin{figure}[htb]
\includegraphics[width=8cm,height=6cm, clip]{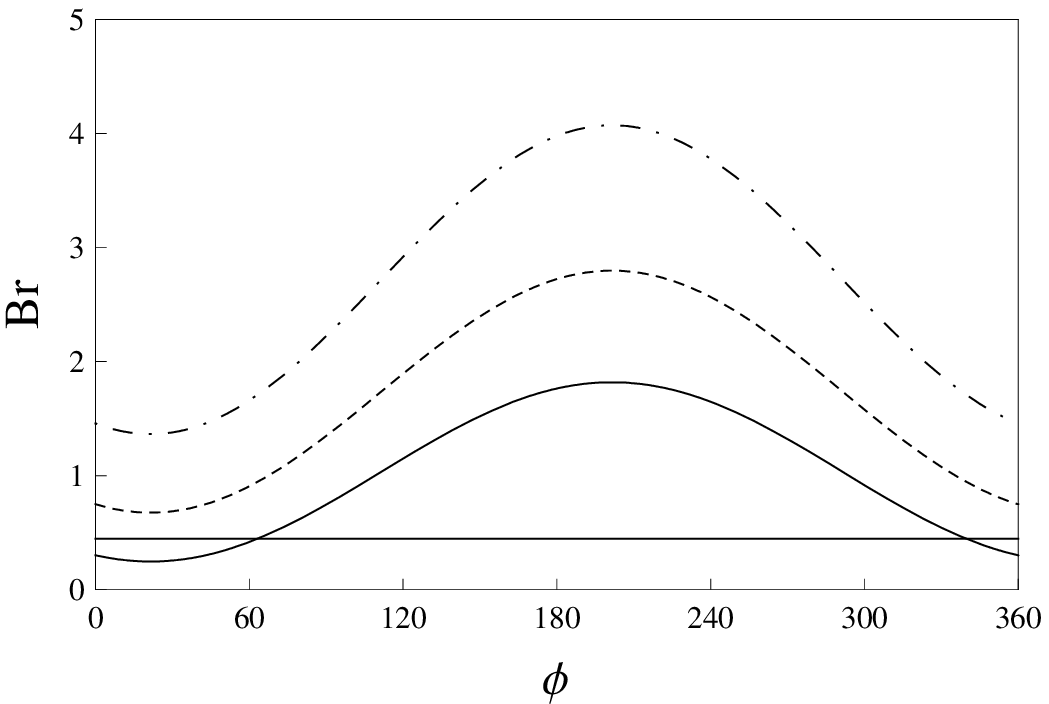}
\hspace{0.2cm}
\includegraphics[width=8cm,height=6cm, clip]{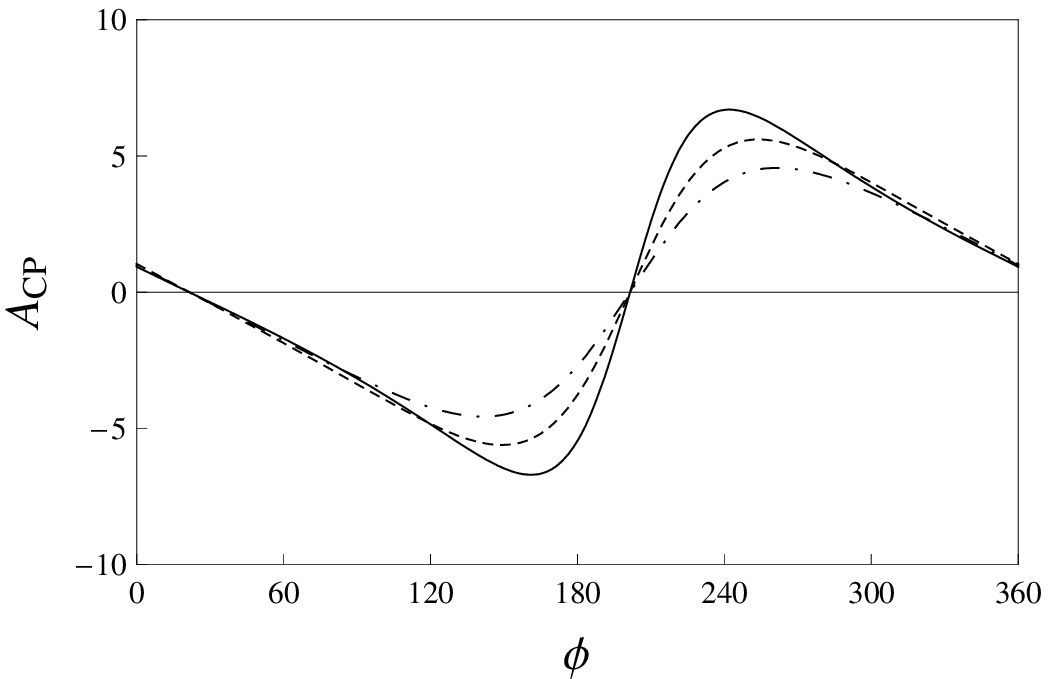}
\caption{Variation of CP averaged branching ratio (\ref{br}) (in
units of $10^{-8}$)  (left panel) and direct CP asymmetry (\ref{cp})
(in \%) (right panel) with the new weak phase $\phi$,  where the
solid, dashed and dot-dashed lines correspond to $m_{t'}=400, 600$
and 800 GeV respectively. The horizontal line in the left panel
represents the SM value. }
\end{figure}

Now we consider another extension of the SM, where the fermion
sector is enlarged by an extra down type singlet quark. Isosinglet
quarks appear in many extensions of the SM like the low energy limit
of the $E_6$ GUT models \cite{e6}. The mixing of this singlet type down quark
with the three SM down type quarks  provides a framework to study
the deviations of the unitarity constraint of the $3 \times 3 $ CKM
matrix. The mixing also induces tree level flavor changing neutral
currents, which can thus substantially modify the SM results. In
this model the $Z$ mediated FCNC interaction is given by \cite{yg}
\be {\cal
L} = \frac{g}{2 \cos \theta_W} [ \bar d_{L \alpha} U_{\alpha \beta}
\gamma^\mu d_{L \beta}]Z_\mu\;, \ee with \be U_{\alpha \beta}=\sum_{i=u,c,t}
V_{\alpha i}^\dagger V_{i \beta}= \delta_{\alpha \beta}-V_{4
\alpha}^* V_{4 \beta}\;, \ee where $\alpha,~\beta$ are generation
indices and  $U$ is the neutral current mixing
matrix for the down quark sector. The non-vanishing component of
$U_{\alpha \beta}$ will lead to the presence of FCNC transitions at
the tree level. The implications of the FCNC mediated $Z$ boson
effect has been extensively studied in the context of $b$ physics
\cite{rm, desh, vives}.

Because of the new interactions the effective Hamiltonian describing
$b \to d \bar s s$ process is given as \cite{desh}, \be {\cal
H}_{eff}^Z= - \frac{G_F}{\sqrt 2} V_{tb}V_{td}^*[ \tilde C_3 O_3 +
\tilde C_7 O_7 + \tilde C_9 O_9]\;, \ee where the four-quark
operators $O_3$, $O_7$ and $O_9$ have the same structure as the SM
QCD and electroweak penguin operators and the new Wilson
coefficients $\tilde C_i$'s at the $M_Z$ scale are given by
\bea \tilde C_3(M_Z) &=& \frac{1}{6} \frac{U_{bd}}{V_{tb}V_{td}^*},\nn\\
\tilde C_7(M_Z) &=& \frac{2}{3} \frac{U_{bd}}{V_{tb}V_{td}^*}
\sin^2 \theta_W,\nn\\
\tilde C_9(M_Z) &=& -\frac{2}{3} \frac{U_{bd}}{V_{tb}V_{td}^*}
(1- \sin^2 \theta_W).
 \eea
These new Wilson coefficients will be evolved from the $M_Z$ scale
to the $m_b$ scale using renormalization group equation \cite{wilson}
as described
earlier. Because of the RG evolution these three Wilson coefficients
generate new set of Wilson coefficients $\tilde C_i (i=3,\cdots,
10)$ at the low energy regime (i.e., at the $m_b$ scale) as presented in
Table-1. Thus, one
can write the new amplitude due to the tree level FCNC mediated $Z$
boson effect in a straight forward manner from Eq. (\ref{amp}) by
replacing $\alpha_{3(3,EW)}$ by $\tilde \alpha_{3(3,EW)}$, where
$\tilde \alpha$'s are related to the new Wilson coefficients $\tilde
C_i(m_b)$'s. In order to see the effect of this  FCNC mediated $Z$
boson effect we have to know the value of the parameter $Z-b-d$
coupling parameter  which can be explicitly written as
$U_{bd}=|U_{bd}|e^{i \phi}$ and the allowed range of
$|U_{bd}|$ is found to be $(2 \times 10^{-4} \leq |U_{bd}|
\leq 1.2 \times 10^{-3})$ \cite{vives}. In Figure-2, we
present the variation of
the CP averaged branching ratio (\ref{br}) with $|U_{bd}|$ and $\phi$
(left panel) and the direct CP asymmetry parameter $A_{CP}$ with $\phi$
(right panel),
where we have used $\sin^2\theta_W=0.231$. From  figure-2 it can be
seen that the branching ratio could be significantly enhanced and
large CP violation could be possible in this
model.

\begin{figure}[htb]
\includegraphics[width=8cm,height=6cm, clip]{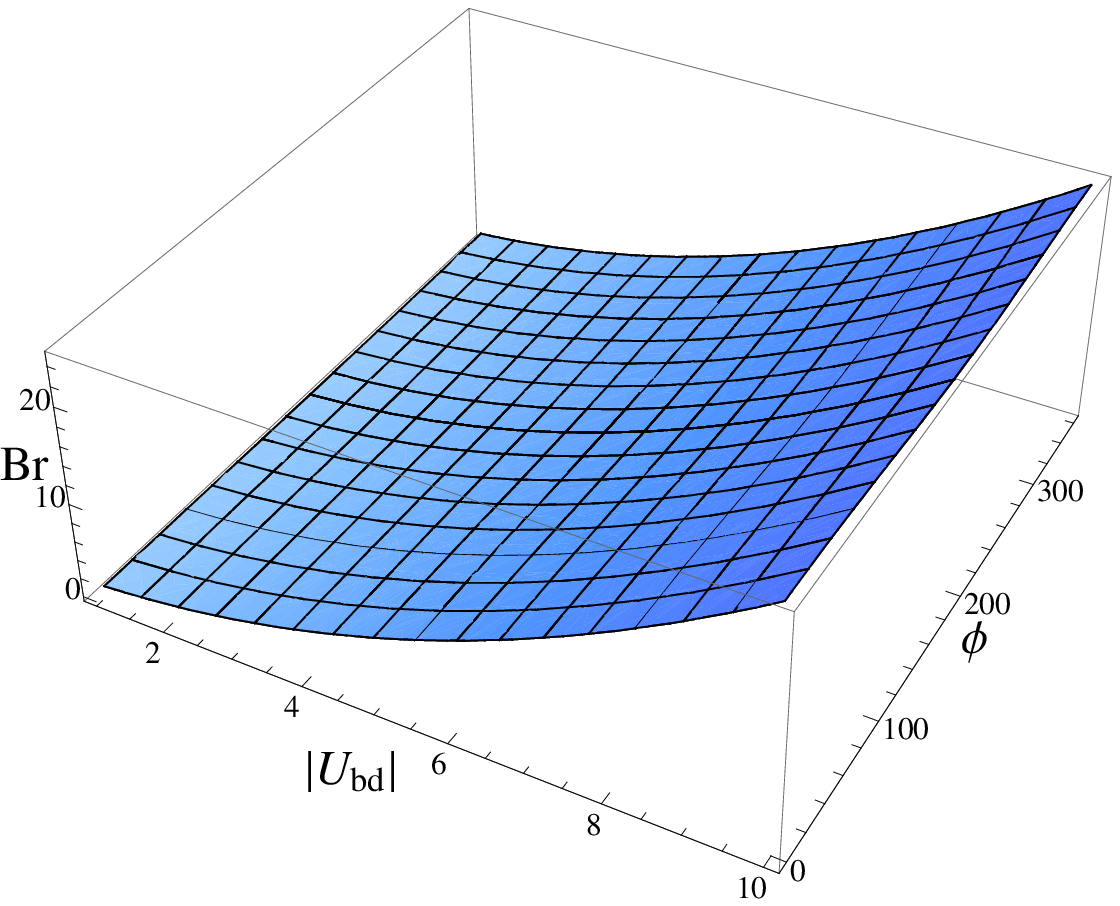}
\hspace{0.2 cm}
\includegraphics[width=8cm,height=6cm, clip]{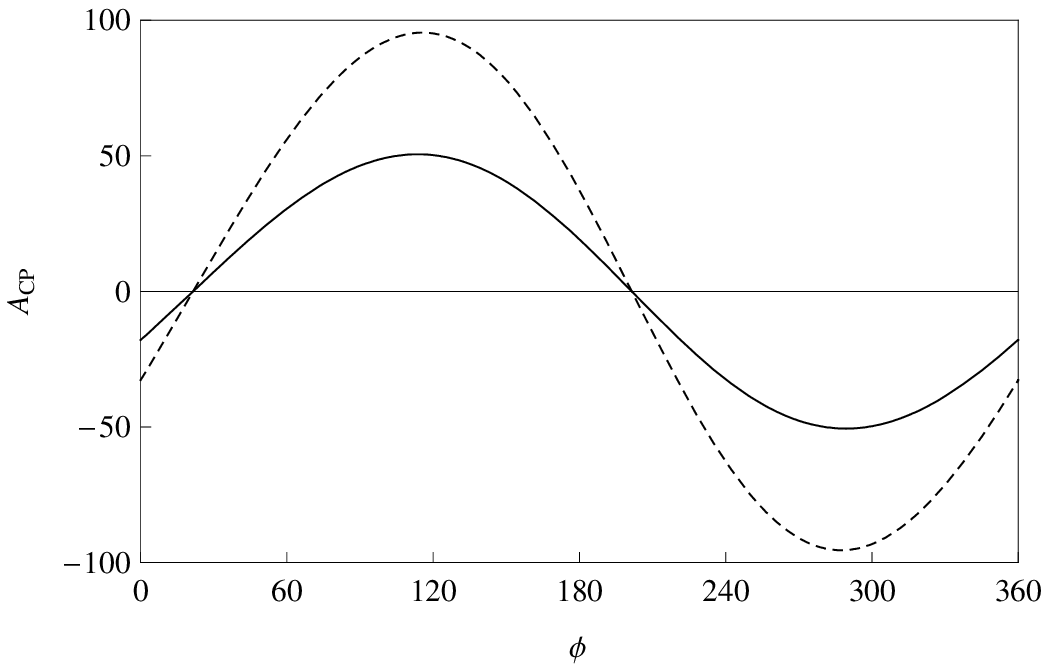}
\caption{Variation of the CP averaged branching ratio (in units of
$10^{-8}$) with $|U_{bd}|$ (in units of $10^{-4}$) and the new weak
phase $\phi$ (left panel) and the variation of  direct CP asymmetry (\ref{cp})
 (in \%)  with the new weak phase $\phi$ (right panel)
where the dashed and solid lines
correspond to $|U_{bd}|=10^{-4}$ and $5 \times 10^{-4}$.}
\end{figure}

Now we consider the effect due to an extra $U(1)'$ gauge boson $Z'$.
The existence of extra $Z'$ boson  is a feature of many models
addressing physics beyond the SM, e.g., models based on extended
gauge groups characterized by additional $U(1)$ factors \cite{zp1}.
Also the new physics models which contain exotic fermions, predict
the existence of additional gauge boson. Flavor mixing can be
induced at the tree level in the up-type and/or down-type quark
sector after diagonalizing their mass matrices. Here again as in the
$Z$ model, FCNCs due to $Z'$ exchange can be induced by mixing among
the SM quarks and the exotic quark which have different $Z'$ quantum
numbers. Here we will consider the model in which the interaction
between the $Z'$ boson and fermions are flavor nonuniversal for left
handed couplings and flavor diagonal for right handed couplings. The
detailed description of the model can be found in Ref. \cite{zp2, zp2a},
where it has been shown that such model can successfully explain the
deviations of $S_{\phi K}$ and $S_{\eta' K}$ from   $S_{\psi K}$ and
also can explain the $B \to \pi K$ puzzle. The search for the extra
$Z'$ boson occupies an important place in the experimental programs
of the Fermilab Tevatron and CERN LHC \cite{zp3}. At such hadron colliders
heavy neural gauge bosons with mass upto around 5 TeV can be
produced and detected via two fermion decays $pp(p \bar p) \to Z' \to l^+ l^-
~(l=e,\mu)$.

The effective Hamiltonian describing the transition $b \to d \bar s
s$ mediated by the $Z'$ boson is given by \cite{zp2} \be {\cal
H}_{eff}^{Z'}= - \frac{4 G_F}{\sqrt 2} V_{tb}V_{td}^* \left [ \left
( \frac{g' M_Z} {g_1 M_{Z'}} \right )^2 \frac{B_{db}^L}{V_{tb}
V_{td}^*} (B_{ss}^L O_9 +B_{ss}^R O_7)\right ]\;, \label{zp} \ee
where $g_1=e/(\sin \theta_W \cos \theta_W)$ and $B_{ij}^{L(R)}$
denote the left (right) handed effective $Z'$ couplings of the
quarks $i$ and $j$ at the weak scale. The diagonal elements are real
due to the hermiticity of the effective Hamiltonian but the off
diagonal elements may contain effective weak phase. Therefore, both
the terms in (\ref{zp}) will have the same weak phase due to
$B_{db}^L$. We can parameterize these coefficients as \be \xi_L=
\left ( \frac{g' M_Z} {g_1 M_{Z'}} \right )^2 \left ( \frac{B_{db}^L
B_{ss}^L}{V_{tb}V_{td}^*}\right )=|\xi_L| e^{i \phi'}\;,~~~ \xi_R=
\left ( \frac{g' M_Z} {g_1 M_{Z'}} \right )^2 \left (\frac{B_{db}^L
B_{ss}^R}{V_{tb} V_{td}^*}\right )=|\xi_R| e^{i \phi'}\;, \ee where
$\phi'=\phi-\beta$, ($\phi$ is the weak phase associated with
$B_{db}^L$).

In order to see the effect of $Z'$ boson, we have to know 
the values of the $\xi_L$ and $\xi_R$
or equivalently $B_{db}^L$ and $B_{ss}^{L,R}$. Assuming only left
handed couplings are present, the bound on FCNC $Z'$ coupling
($B_{db}^L)$ from $B^0 - \bar B^0$ mass difference has been obtained
in Ref. \cite{desh1} as \be y|{\rm Re}(B_{db}^L)^2| < 5 \times
10^{-8},~~~~ y|{\rm Im }(B_{db}^L)^2| < 5 \times 10^{-8}\;, \ee
where $ y=(g' M_Z /g_1 M_{Z'})^2$. Generally one expects $g'/g_1
\sim 1 $, if both the $U(1)$ gauge groups have the same origin from
some grand unified theories, $M_Z/M_{Z'} \sim 0.1 $ for a TeV scale
neutral $Z'$ boson, which yields $y \sim 10^{-2}$. However in Ref. \cite{desh1}
assuming a small mixing between $Z-Z'$ bosons the value of $y$ is
taken as $y \sim 10^{-3}$. Using $y \sim 10^{-2}$, one can obtain a
more stringent bound on $|B_{db}^L| < 10^{-3}$. 
It has been shown in \cite{zp2a} that the mass difference
of $B_s - \bar B_s $ mixing can be explained if $|B_{sb}^L| \sim
|V_{tb} V_{ts}^*| $. Similarly, the CP asymmetry anomaly in  $B \to
\phi K, \pi K $   can be resolved if $|B_{sb}^L B_{ss}^{L,R}| \sim
|V_{tb} V_{ts}^*|$. From these two relations one can obtain
$|B_{ss}^L| \sim 1 $. Thus, it is expected that $\xi_{L,R} \sim
10^{-3}$. However, in this analysis we vary  their values within the
range $(0.01-0.001)$.

After having an idea about the magnitudes of these new coefficients
which are at the $M_Z$ scale, we  now evolve them to the $b$ scale
using renormalization group equation \cite{wilson}.
The new Wilson coefficients at the $m_b$ scale are presented in Table-1.
Using the values of these
coefficients at $b$ scale we can analogously obtain the new
contribution to the transition amplitude as done in the case of $Z$
boson. Now using $|\xi_L|=|\xi_R|=\xi$, in figure-3, we show the
variation of the  CP averaged branching ratio with $\xi$ and the new
weak phase $\phi$ (left panel) and the direct CP violation with
$\phi$ (right panel). In this case also one can have a significant
enhancement in the branching ratio for large $\xi$, or in other
words for a lighter $Z'$ boson. Furthermore, the observation
of this mode could in turn help us to constrain the $Z'$ mass.

\begin{figure}[htb]
\includegraphics[width=8cm,height=6cm, clip]{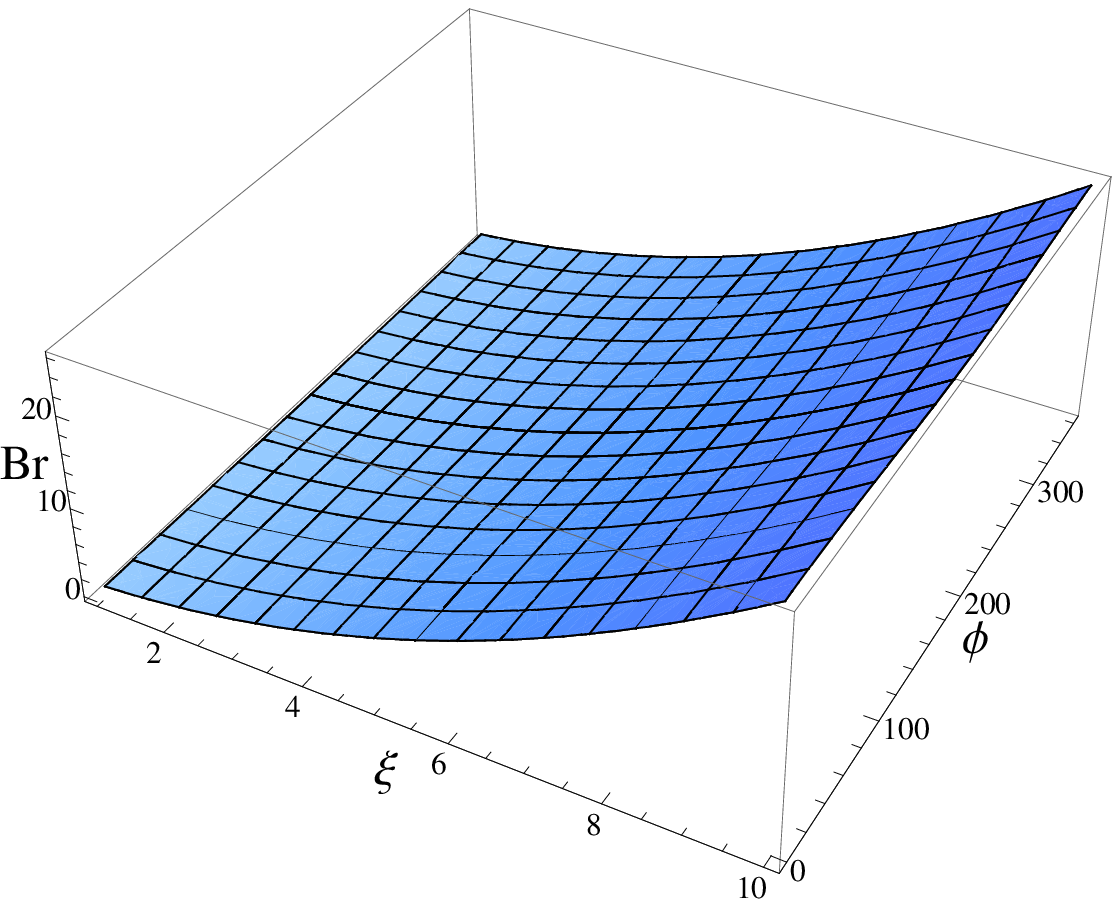}
\hspace{0.2 cm}
\includegraphics[width=8cm,height=6cm, clip]{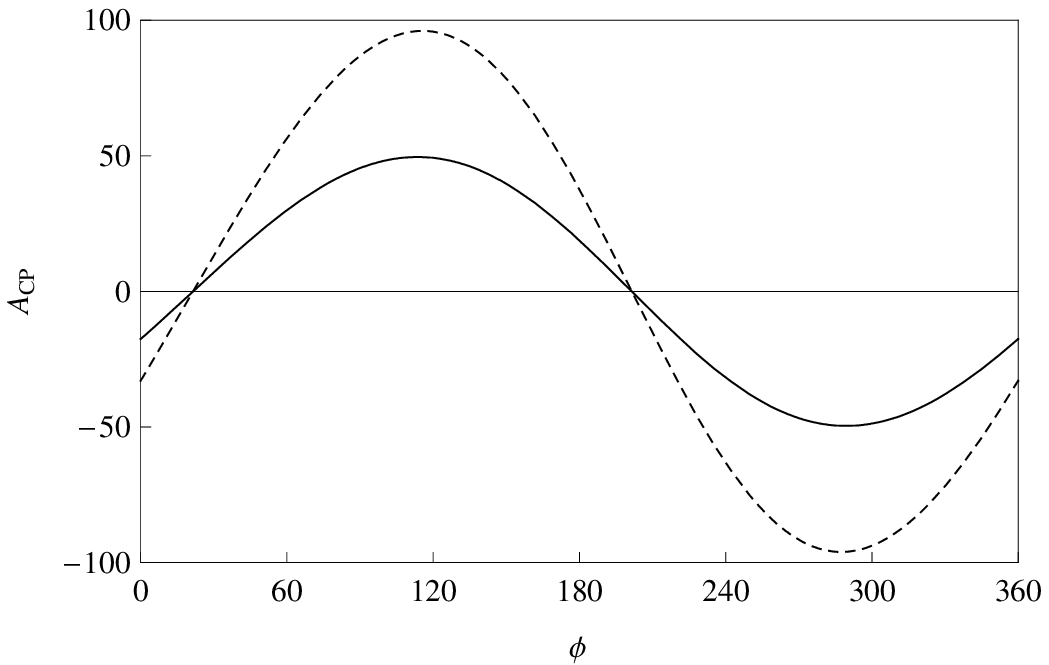}
\caption{Variation of the CP averaged branching ratio (in units of
$10^{-8}$) with $\xi$ (in units of $10^{-3}$) and the new weak phase
$\phi$ (left panel) and the variation of  direct CP asymmetry (\ref{cp})
 (in \%)  with the new weak phase $\phi$ (right panel)
where the dashed and solid lines
correspond to $\xi=10^{-3}$ and $5 \times 10^{-3}$.}
\end{figure}

In this paper, we have studied the $B^- \to \phi \pi^-$ decay mode
in the standard model and in some beyond the standard model
scenarios. This is a pure penguin rare decay process and proceeds
through the quark level transition $ b \to d  \bar s s$,  which
occurs at the one loop level and is therefore expected to be highly
suppressed in the SM. The SM prediction of its branching ratio is
$\sim {\cal O} (10^{-9})$ which is below the experimental upper
limit of ${\cal O} (10^{-7})$. We have analysed this decay mode in
the fourth quark generation model and in the FCNC mediated $Z$ and
$Z'$ models.  In the fourth quark generation model, we find that the
branching ratio enhances from its SM value, with
the increasing $m_{t'}$ and it can have a value of $\sim{\cal O}
(10^{-8})$. In the $Z$ and $Z'$ models,  the branching ratio can be
significantly enhanced for sizable new physics couplings $|U_{bd}|$
and $\xi$. In these cases it can reach up to  ${\cal O} (10^{-7})$
level but still within the experimental upper limit. Furthermore, it
is found that large direct CP violation could be possible in this
decay mode in the presence of above mentioned new physics models.
Thus, if this mode could be observed in the upcoming LHCb experiment
it will provide a clear signal of new physics and also can be used
to constrain the parameter space of various new physics models.
However, it would not be possible to  distinguish between these new
physics models considering this mode alone.

{\acknowledgments}
BM and AKG would like to thank Council of Scientific and Industrial
Research, Government of India for financial support. The work of RM
was partly supported by Department of Science and Technology,
Government of India through grant No. SR/S2/HEP-04/2005.


\begin{thebibliography} {99}
\bibitem{S} E. Barberio et al., [HFAG Collaboration], arXiv:
0704.3575[hep-ex].

\bibitem{soni} E. Lunghi and A. Soni, JHEP {\bf 09}, 053, (2007)
(arXiv:0707.0212[hep-ph]).

\bibitem{prd74} B. Aubert {\it et al.}, [Babar Collaboration],
Phys. Rev. D {\bf 74}, 011102 (2006).

\bibitem{phism} D. Du, H. Gong, J. Sun, D. Yang and G. Zhu, Phys.
Rev. D {\bf  65}, 094025 (2002); {\it ibid.} {\bf  66}, 079904 (2002),
 Erratum; R. Aleksan,
P. -F. Giraud, V. Morenas, O. Pene and A. S. Safir, Phys. Rev. D {\bf
 67}, 094019 (2003); B. Melic, Phys. Rev. D {\bf 59}, 074005
(1999); A. Ali, G. Kramer and C. D. Lu, Phys. Rev. D {\bf  58},
094009 (1998); R. Fleischer, Phys. Lett. B {\bf 321}, 259 (1994); D.
S. Du and Z. Z. Xing, Phys. Lett. B {\bf 312}, 199 (1993).

\bibitem{phi} S. Bar-Shalom, G. Eilam and Y. D. Yang, Phys.
Rev. D {\bf 67}, 014007 (2003); J. F. Cheng and C. S. Huang, Phys.
Lett. B {\bf 554}, 155 (2003); A. K. Giri and R. Mohanta, Phys.
Lett. B {\bf 594}, 196 (2004); {\it ibid.} {\bf 660}, 376 (2008); J.
F. Cheng, Y. N. Gao, C. S. Huang and X. H. Wu, Phys. Lett. B {\bf
647}, 413 (2007).

\bibitem{qcd} M. Beneke and M. Neubert, Nucl. Phys. B {\bf 675},
333 (2003); M. Beneke, G. Buchalla, M. Neubert and C. T. Sachradja,
Nucl. Phys. B {\bf 606}, 245 (2001).

\bibitem{pdg} Particle Data Group: W. M. Yao et al., J. Phys.
{\bf G 33}, 1 (2006).

\bibitem{4gen} W. -S. Hou, A. Soni and H. Steger, Phys. Lett. B {\bf
192}, 441 (1987); W. S. Hou, R. S. Willey and A. Soni, Phys. Rev.
Lett. {\bf 58}, 1608 (1987).

\bibitem{4thgen} W. -S. Hou, M. Nagashima, G. Raz and A.
Soddu, JHEP {\bf 09}, 012 (2006);  W. -S. Hou, M. Nagashima and A.
Soddu, Phys. Rev. Lett {\bf 95}, 141601 (2005);  W. -S. Hou, H. -N.
Li, S. Mishima and M. Nagashima, Phys. Rev. Lett. {\bf 98}, 131801
(2007);
 T. M. Aliev, A.
Ozpineci and M. Savci, Euro. Phys. J. C {\bf 29}, 265 (2003);
 L. Solmaz, Phys. Rev. D {\bf 69}, 015003 (2004);
T. Hasuike, T. Hattori, T. Hayashi and S. Wakaizumi, Phys. Rev. D {\bf
41}, 1691 (1990).


\bibitem{hou2} A. Arhrib and W. -S. Hou, Euro. Phys. J. C {\bf 27}, 555
(2003).

\bibitem{lim} T. Inami and C. S. Lim, Prog. Theor. Phys. {\bf 65},
297 (1981); {\it ibid} {\bf 65}, 1772E (1981).


\bibitem{wilson} G. Buchalla, A. J. Buras and M. Lautenbacher,
Rev. Mod. Phys. {\bf 68}, 1125 (1996).

\bibitem{cdf} CDF collaboration, http://www-cdf.fnal.gov/physics/new/top/2008/tprop/Tprime2.3/cdf92
34\_tprime\_23\_pub.pdf

\bibitem{soni08} A. Soni, A. K. Alok, A. Giri, R Mohanta and S.
Nandi, arxiv:0807.1971 (hep-ph).

\bibitem{e6}  M. Bando and T. Kugo, Prog. Theor. Phys. {\bf 101},
1313 (1999); M. Bando, T. Kugo and K. Yoshioka, Prog. Theor.
Phys. {\bf 104}, 211 (2000).

\bibitem{yg} Y. Grossman, Y. Nir and R. Rattazzi, in {\it Heavy Flavours
II}, edited by A. J. Buras and M. Lindner (World Scientific Singapore, 1998),
p.755.

\bibitem{rm} M. Gronau and D. London, Phys. Rev. {\bf D 55},
2845 (2000); A. K. Giri and R. Mohanta, Phys. Rev. D {\bf 68},
014020 (2003); {\it ibid} {\bf 69}, 014008 (2004); Mod. Phys. Lett.
A {\bf 19}, 1903 (2004); JHEP {\bf 11}, 084 (2004); Euro. Phys. J. C
{\bf 45}, 151 (2006);  R. Mohanta, Phys. Rev. D {\bf 71}, 114013
(2005).

\bibitem{desh} D. Atwood and G. Hiller, hep-ph/0307251;
N. G. Deshpande and D. K. Ghosh, Phys. Lett. B {\bf
593}, 135 (2004).

\bibitem{vives}  G. Barenboim, F. J. Botella and O. Vives, Phys.
Rev. D {\bf 64}, 015007 (2001); Nucl. Phys. B {\bf 613}, 285 (2001).


\bibitem{zp1} P. Langacker and. M. Pl\"umacher, Phys. Rev. D {\bf 62},
 013006 (2000); P. Langacker, hep-ph/0308033.

\bibitem{zp2} V. Barger, C. W. Chiang, P. Langacker and H. S. Lee,
Phys. Lett. B {\bf 580}, 186 (2004); {\it ibid.} {\bf 598}, 218
(2004).

\bibitem{zp2a}  V. Barger, C. W. Chiang, J. Jiang and  P. Langacker,
 Phys. Lett. B {\bf
596}, 229 (2004).

\bibitem{zp3} T. G. Rizzo, hep-ph/0610104; P. Langacker,
arxiv:0801.1345 (hep-ph).

\bibitem{desh1} C. W. Chiang, N. G. Deshpande and J. Jiang, JHEP08,
075 (2006).



\end{thebibliography}
\end{document}